\documentclass[fleqn,usenatbib]{mnras}

\usepackage[T1]{fontenc}

\DeclareRobustCommand{\VAN}[3]{#2}
\let\VANthebibliography\thebibliography
\def\thebibliography{\DeclareRobustCommand{\VAN}[3]{##3}\VANthebibliography}

\usepackage{graphicx}	
\usepackage{amsmath}	
\usepackage{amssymb}	
\usepackage{subcaption}
\usepackage[normalem]{ulem}
\usepackage[dvipsnames]{xcolor}

\usepackage{newtxtext,newtxmath}

\renewcommand{\Vec}[1]{\mathbfit{#1}}
\newcommand{\dd}{{\rm d}}

\title[GW scaling with vortex number]{Persistent gravitational radiation from glitching pulsars. II. Updated scaling with vortex number}

\author[Cheunchitra et al.]{
T.~Cheunchitra,$^{1, 2}$\thanks{E-mail: tcheunchitra@student.unimelb.edu}
A.~Melatos,$^{1,2}$ J.~B.~Carlin,$^{1,2}$ and G.~Howitt$^{3,4}$
\\
$^{1}$School of Physics, University of Melbourne, Parkville, Victoria 3010, Australia\\
$^{2}$OzGrav, Australian Research Council Centre of Excellence for Gravitational Wave Discovery, University of Melbourne, Parkville, Victoria 3010, Australia\\
$^{3}$Computational Biology Program, Peter MacCallum Cancer Centre, Melbourne, Victoria 3000, Australia \\
$^{4}$Sir Peter MacCallum Department of Oncology, University of Melbourne, Parkville, Victoria 3010, Australia 
}

\date{Accepted XXX. Received YYY; in original form ZZZ}

\pubyear{2023}

\begin{document}
\label{firstpage}
\pagerange{\pageref{firstpage}--\pageref{lastpage}}
\maketitle

\begin{abstract}
    Superfluid vortices pinned to nuclear lattice sites or magnetic flux tubes in a neutron star evolve abruptly through a sequence of metastable spatial configurations, punctuated by unpinning avalanches associated with rotational glitches, as the stellar crust spins down electromagnetically.
    The metastable configurations are approximately but not exactly axisymmetric, causing the emission of persistent, quasimonochromatic, current quadrupole gravitational radiation. 
    The characteristic gravitational wave strain $h_0$ as a function of the spin frequency $f$ and distance $D$ from the Earth is bounded above by $h_0 = 1.2\substack{+1.3 \\ -0.9} \times 10^{-32} (f/30\;{\rm Hz})^{2.5} (D/1\;{\rm kpc})^{-1}$, corresponding to a Poissonian spatial configuration (equal probability per unit area, i.e. zero inter-vortex repulsion), and bounded below by $h_0 = 1.8\substack{+2.0 \\ -1.5} \times 10^{-50} (f/30\;{\rm Hz})^{1.5} (D/1\;{\rm kpc})^{-1}$, corresponding to a regular array (periodic separation, i.e.\ maximum inter-vortex repulsion). 
    N-body point vortex simulations predict an intermediate scaling, $h_0 = 7.3\substack{+7.9 \\ -5.4} \times 10^{-42} (f/30\;{\rm Hz})^{1.9} (D/1\;{\rm kpc})^{-1}$, which reflects a balance between the randomizing but spatially correlated action of superfluid vortex avalanches and the regularizing action of inter-vortex repulsion. 
    The scaling is calibrated by conducting simulations with ${N_{\rm v}} \leq 5\times10^3$ vortices and extrapolated to the astrophysical regime ${N_{\rm v}} \sim 10^{17} (f/30\;{\rm Hz})$. 
    The scaling is provisional, pending future computational advances to raise ${N_{\rm v}}$ and include three-dimensional effects such as vortex tension and turbulence.
\end{abstract}




\section{Introduction}
    
    Pulsars that undergo spin-up glitches are predicted to emit gravitational waves at amplitudes that may be detectable by long-baseline interferometers like the Laser Interferometer Gravitational-wave Observatory (LIGO). 
    The predicted signals fall into two categories: 
    (i) bursts lasting $\lesssim 1 \, {\rm s}$, which occur during spin-up events due to the excitation of stellar oscillation modes \citep{AnderssonComer2001, Sideryetal2010, Hoetal2020, Andersson2021, YimJones2022} or the reorganization of the superfluid velocity field inside the star \citep{WarszawskiMelatos2012, DeLilloEtAl2022}; or 
    (ii) quasimonochromatic signals lasting $\lesssim {\rm weeks}$, which occur after the spin-up event due to relaxation processes such as Ekman circulation \citep{vanEysdenMelatos2008, Bennettetal2010, Singh2017} or subsidence of a transient mountain \citep{YimJones2020, MoraguesEtAl2023}, for which there is evidence in radio timing data \citep{vanEysdenMelatos2010}. 
    Both categories have been the target of LIGO searches \citep{Abadieetal2011, Abbottetal2021, Prestegard2016} and the development of search algorithms \citep{Thraneetal2011, Prixetal2011, Thraneetal2015, KeitelAshton2018, Milleretal2018, Oliveretal2019, Macquetetal2021, LopezEtAl2022, MoraguesEtAl2023}.
    
    Superfluid vortex avalanches are a standard theoretical explanation for neutron star glitches \citep{AndersonItoh1975, WarszawskiMelatos2011} although there are other viable mechanisms as well, such as starquakes \citep{Middleditchetal2006, ChugunovHorowitz2010, KerinMelatos2022, MoralesHorowitz2023} and hydrodynamical instabilities \citep{Anderssonetal2003, Peraltaetal2006, GlampedakisAndersson2009}; see \citet{HaskellMelatos2015}, \citet{AntonopoulouEtAl2022}, \citet{Zhou2022}, and \citet{AntonelliEtAl2022} for recent reviews. 
    In a superfluid vortex avalanche, a large number of vortices collectively and spontaneously unpin and migrate outward, transferring angular momentum to the crust before they repin or annihilate at the crust. 
    The vortices transition impulsively from one metastable configuration before the avalanche to another after the avalanche, i.e.\ the time-scale on which the vortices adjust their configuration is much shorter than the time-scale between avalanches. 
    The configuration between avalanches does not change due to pinning \citep{LinkEpstein1991, LinkEpsteinBaym1993, DonatiPizzochero2006, AvogadroEtAl2008, SevesoEtAl2016}.
    None of the metastable configurations is the stable, ground-state configuration, i.e. the regular, triangular, Abrikosov array; they are all slightly nonaxisymmetric perturbations to the ground-state\footnote{Superfluid turbulence on macroscopic and microscopic scales complicates the three-dimensional picture further and raises formidable computational challenges \citep{Greenstein1970, PeraltaEtAl2005, Peraltaetal2006, AnderssonEtAl2007, PeraltaMelatos2009, MelatosPeralta2010, MongioviEtAl2017, DrummondMelatos2017, DrummondMelatos2018, KhomenkoEtAl2019, HaskellEtAl2020, ThongEtAl2023, LevinLink2023}.}.
    Interestingly, this means that the superfluid velocity field is nonaxisymmetric at all times, not just during or immediately before or after a glitch, which in turn means that the star always has a nonzero, time-dependent current quadrupole moment \citep{Melatosetal2015, HaskellEtAl2022}. 
    The nonaxisymmetry is small, but it may be large enough to produce gravitational wave signals detectable using long integrations of data from a detector such as LIGO in the medium-term future. 
    Calculating this current quadrupole signal is the goal of this paper.
    
    Previous efforts to calculate the current quadrupole moment have used Gross-Pitaevskii simulations to evolve the pinned vortex array during and between unpinning events \citep{WarszawskiMelatos2012, Melatosetal2015}. 
    These calculations are limited by computational cost to ${\sim}10^2$ vortices. 
    This is much less than the  ${N_{\rm v}} \sim 10^{15} (\Omega / 1\,{\rm rad\,s^{-1}})$ vortices in a typical neutron star. 
    A natural next step is to calculate the current quadrupole moment analytically. 
    However, this faces a fundamental challenge: there is no general analytic theory of far-from-equilibrium systems like vortex avalanches, which exhibit self-organized criticality and long-range correlations between widely separated vortices, not just nearest neighbors \citep{Jensen1998, Aschwanden2011}.
    This fundamental problem lies outside the scope of this paper.
    Instead, we estimate how the current quadrupole moment scales with ${N_{\rm v}}$ for vortex distributions that approximate plausibly the realistic situation in a neutron star and can be analysed analytically.
    We test the results against an N-body code \citep{Howittetal2020, Howittetal2022}, which simulates avalanches containing up to ${\sim}10^4$ vortices and therefore extends the baseline over which the ${N_{\rm v}}$ scaling can be verified beyond the Gross-Pitaevskii regime --- although, not as far as ${N_{\rm v}} \sim 10^{15} (\Omega / 1\,{\rm rad\,s^{-1}})$ as in a realistic neutron star.
    The results are compared with the hydrodynamic limit calculated by \citet{HaskellEtAl2022}.
    
    The paper is structured as follows. 
    In Section 2, we summarize the physics of a vortex avalanche. 
    In Section 3, we relate the gravitational wave strain to the current quadrupole moment for an arbitrary array of rectilinear vortices. 
    In Section 4, we use the N-body simulations to calculate the current quadrupole moment and hence the gravitational wave strain as functions of ${N_{\rm v}}$ in the regime $5\times10^2 \leq {N_{\rm v}} \leq 5\times10^3$. 
    In Section 5, we consider two idealized vortex configurations as upper and lower limits, namely a uniform Poisson point process and a regular periodic array respectively, motivated by the output of the N-body simulations. 
    The current quadrupole moment and gravitational wave strain are calculated analytically for both configurations as functions of ${N_{\rm v}}$, so that the N-body results for $5\times10^2 \leq {N_{\rm v}} \leq 5\times10^3$ lie between them. 
    In Section 6, we extrapolate the N-body and analytic scalings obtained in Sections 4 and 5 to realistic values of ${N_{\rm v}}$ for neutron stars. 
    We comment briefly on the astrophysical implications and uncertainties, including the possibility of superfluid turbulence, in Section 7.

\section{Superfluid Vortex Avalanches}
    \label{sec:SuperfluidVortexAvalanches}
    
    The neutron superfluid inside a neutron star attempts to rotate uniformly and match the angular velocity $\Omega$ of the crust by forming vortices, each of which carries a quantum of circulation $\kappa = h/2m_n$, where $m_n$ is the mass of the neutron, and $h$ is Planck's constant. 
    The number of vortices, ${N_{\rm v}}$, is therefore determined by the total circulation, $2\pi R_\ast^2 \Omega$, where $R_\ast$ is the radius of the neutron condensate \citep{AndersonItoh1975}. 
    As the crust decelerates, a fictitious Magnus force pulls vortices radially outward. 
    However, vortices are embedded in a nuclear lattice and an array of magnetic flux tubes, to which they pin \citep{Srinivasanetal1990, PethickSmith2001, SevesoEtAl2016, DrummondMelatos2017, DrummondMelatos2018, ThongEtAl2023}. 
    When the Magnus force exceeds a threshold, one or more vortices unpin at random, through single- or multi-site breakaway \citep{LinkEpstein1991, LinkEpsteinBaym1993, LinkLevin2022}, and knock-on their neighbours through proximity or acoustic mechanisms \citep{Warszawskietal2012}. 
    Knock-on triggers a runaway process, in which anywhere from one to $N_{\rm v}$ vortices unpin --- that is, an avalanche. 
    The associated transfer of angular momentum to the crust produces a spin-up glitch.
    
    Gross-Pitaevskii simulations of vortex avalanches are used to study pulsar glitches. 
    They demonstrate key aspects of the physics, such as the operation of proximity and acoustic knock-on, and yield estimates of the probability density functions of key stochastic variables, e.g. avalanche waiting times (typically an exponential), avalanche sizes (typically a power law), radial vortex displacements (typically a Gaussian), and so on \citep{WarszawskiMelatos2011, Melatosetal2015}.
    However, computational cost limits such simulations to ${N_{\rm v}} \lesssim 10^2$ vortices.
    
    Recently, \citet{Howittetal2020} developed N-body simulations which show glitch-like behaviour for larger systems with ${N_{\rm v}} \leq 5\times10^3$. The N-body simulations produce avalanches whose internal structure (e.g. wedge-like shape) and externally observable stochastic properties (e.g. waiting times and sizes) resemble qualitatively those produced by Gross-Pitaevskii simulations \citep{Howittetal2020, Howittetal2022}. 
    Furthermore, the N-body simulations predict scalings between observables and control parameters (e.g. spin-down torque, pinning site separation, pinning potential), which agree with those produced by Gross-Pitaevskii simulations; compare, for example, Table 2 in \citet{Howittetal2020} with Table 9 in \citet{WarszawskiMelatos2011}. 
    Encouragingly, this occurs even though the N-body simulations neglect certain physics that is present in Gross-Pitaevskii simulations (e.g. acoustic knock-on) and include certain physics that is absent from Gross-Pitaevskii simulations (e.g. mutual friction).
    
\begin{figure}
    	\includegraphics[width=\columnwidth]{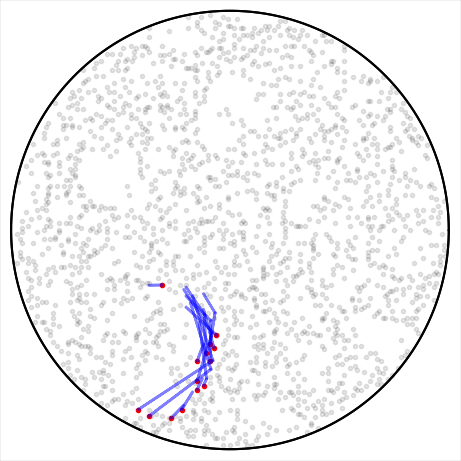}
        \caption{Snapshot of the vortex distribution extracted from an N-body simulation of a pinned, decelerating superfluid with the code developed by \citet{Howittetal2020}. The snapshot is displayed in the frame corotating with the pinning sites. 
        The vortex distribution is shown after the avalanche ends. Vortices that move in the avalanche are drawn as red dots, with a blue tracer showing their discretized path throughout the avalanche. 
        Vortices that do not move are drawn as grey dots. 
        The simulation is initialized by drawing positions of $2\times10^3$ vortices from a spatially uniform distribution within container radius $R_\ast=10.0$. 
        Pinning sites are arranged in a rectangular lattice (lattice size $a_{\rm pin}=0.1$, site width $\xi=0.02$) (all lengths measured in simulation units; see Appendix \ref{sec:AppxSimParam}). 
        The snapshot is taken at simulation time-step $T=826$ (midway between the 34\textsuperscript{th} and 35\textsuperscript{th} glitches) and the track shows movement starting from $T=798$ (midway between the 33\textsuperscript{th} and 34\textsuperscript{th} glitches; time in simulation units). Control parameters are recorded in Appendix \ref{sec:AppxSimParam}; see also Sections 2 and 3 in \citet{Howittetal2020}. 
        Vortices often annihilate at the boundary during an avalanche, although none does so in this figure.}
        \label{fig:GH2020SimDisplay}
    \end{figure}

    Figure \ref{fig:GH2020SimDisplay} depicts a snapshot of a typical vortex avalanche generated by the N-body code. 
    The snapshot is taken after the avalanche ends and is displayed in the frame corotating with the pinning sites. 
    Vortices that move during the avalanche are marked in red, with blue tracers tracking their discretized paths during the avalanche. 
    Vortices that do not move before, during, or after the avalanche are drawn in grey. 
    The avalanche is triggered by a single unpinning event at the top left of the group of red dots, slightly displaced from the others, which itself triggers the rest of the avalanche by proximity knock-on. 
    The unpinned vortices occupy a narrow ``wedge'', which curves clockwise to the bottom left in response to the condensate's local departure from corotation and the drag due to mutual friction, which corresponds to rotation through a dissipation angle $\phi$; see Section 2 in \citet{Howittetal2020}. 
    The simulation is initialized with ${N_{\rm v}} = 2000$ vortices. 
    A total of 21 vortices (out of the ${N_{\rm v}} = 1838$ vortices remaining before the avalanche) move during the avalanche, which is the 34th in a sequence starting from a uniform distribution. 
    The local inhomogeneities in the distribution of grey vortices are caused by successive avalanches; many of the grey vortices participate in one of the 33 avalanches leading up to the one depicted in Fig. \ref{fig:GH2020SimDisplay}. 
    Time-lapse movies of avalanches in action can be viewed by the interested reader in fig. 5 (strong mutual friction) and fig. 9 (weak mutual friction) of \citet{Howittetal2020}.
    
\section{Gravitational Radiation}
    \label{sec:GravitationalRadiation}
    Figure \ref{fig:GH2020SimDisplay} makes it clear that pinned vortices are not arranged axisymmetrically between (let alone during) avalanches. 
    The metastable, pinned configuration is clustered and features filaments interspersed with voids, even though there is no preferred axis overall. 
    The sizes and locations of the nonaxisymmetric features are a stochastic outcome of the avalanche history. 
    Indeed, the system exhibits many of the features of self-organized critical systems, such as long-range spatial correlations and hysteresis \citep{Jensen1998, Aschwandenetal2016, AschwandenEtAl2018}. 
    In this section, we calculate the gravitational wave signal expected from the nonaxisymmetric superfluid velocity field generated by the nonaxisymmetric vortex array. 
    The calculation divides into two parts: writing the metric perturbation in terms of the current quadrupole moment (Section 3.1), and calculating the current quadrupole moment as a function of the vortex positions from the simulation output (Section 3.2).
    
    \subsection{Characteristic wave strain}
        \label{CharWaveStrain}
        In the transverse-traceless gauge, the far-field metric perturbation generated by a superposition of time-varying current multipole moments $S^{lm}$ is given by 
        \begin{equation}
            h^{\rm TT}_{jk} = \frac{G}{Dc^5} \sum_{l = 2}^{\infty} \sum_{m = -l}^{l} T^{{\rm B}2, lm}_{jk} \frac{\partial^l S^{lm}}{\partial t^l}
            \label{eq:charwavestrain}
        \end{equation}
        \citep{Thorne1980, WarszawskiMelatos2012, DeLilloEtAl2022}, where $t$ is the retarded time, $D$ is the distance from the source to the observer, $G$ is the Newtonian constant of gravitation, and $c$ is the speed of light in vacuum. 
        $T_{jk}^{{\rm B2}, lm}$ is the $(l,m)$ angular beam pattern, a function of the observer's orientation relative to the source, defined in equation (2.30f) in \citet{Thorne1980}.
        In the Newtonian limit, the current multipole moments simplify to \citep{Thorne1980}
        \begin{align}
            S^{lm} =&\; -\frac{32\pi}{(2l+1)!!} \left[ \frac{l+2}{2l(l-1)(l+1)} \right]^{1/2} \nonumber \\ 
            &\times \int \dd^3 \Vec{x}\, Y^\ast_{lm} r^l \Vec{x} \cdot \nabla \times (\rho \Vec{v})~, 
            \label{eq:currentmultipole}
        \end{align}
        where $Y^\ast_{lm}$ is the complex conjugate of the scalar spherical harmonic of order $(l, m)$, $\rho$ is the mass density, $\Vec{x}$ is the position vector, and $\Vec{v}$ is the velocity field. 
        The integral is performed over the entire source volume.
    
    \subsection{Current quadrupole moment}
        \label{sec:CurrentQuadMoment}
        In this paper, we assume that the superfluid flows incompressibly. 
        Thus, $\rho$ is uniform, the mass multipole moments vanish, and the current multipole moments are proportional to the vorticity $\nabla \times \Vec{v}$. 
        We also assume for simplicity that the vortices are rectilinear, even though possible arguments exist for tangled vorticity in neutron star interiors due to Kolmogorov-like, hydrodynamic turbulence \citep{Greenstein1970, PeraltaEtAl2005, Peraltaetal2006, PeraltaMelatos2009, MelatosPeralta2010, KhomenkoEtAl2019}, Kelvin-wave instabilities like the Donnelly-Glaberson instability \citep{GlabersonJohnsonOstermeier1974, Donnelly1991, AnderssonEtAl2007, MongioviEtAl2017, HaskellEtAl2020, LevinLink2023}, and magnetic flux tube pinning \citep{DrummondMelatos2017, DrummondMelatos2018, ThongEtAl2023}. 
        
        For rectilinear vortices, the leading order contribution to equations (\ref{eq:charwavestrain}) and (\ref{eq:currentmultipole}) are $l = 2, m = \pm 1$. 
        In cylindrical polar coordinates $(R, \phi, z)$, the corresponding spherical harmonic is 
        \begin{equation}
            Y_{2, \pm 1} = \mp \left(\frac{15}{8\pi}\right)^{1/2} \frac{Rz}{R^2 + z^2} e^{\pm {\rm i} \phi} ~. \label{eq:Y21}
        \end{equation}
        Substituting equation (\ref{eq:Y21}) into equation (\ref{eq:currentmultipole}) gives
        \begin{align}
            S^{2, \pm 1} =&\; \left(\frac{512\pi}{405}\right)^{1/2} \int_0^{2\pi} \dd \phi~\cos \phi \nonumber \\
            &\times \int_{0}^{R_\ast} \dd R~R^2 \left( R_\ast^2 - R^2 \right)^{3/2} \left[ \nabla \times (\rho \Vec{v}) \right]_z~, \label{eq:S21}
        \end{align}
        where only the $z$-component of vorticity $\left[ \nabla \times (\rho \Vec{v}) \right]_z$ contributes to the current quadrupole due to the rectilinearity of vortex lines. 
        Each vortex generates a vorticity field
        \begin{equation}
            \nabla \times \Vec{v} = \frac{\kappa}{R}\delta(R-R_i)\delta(\phi-\phi_i)\Vec{e}_z \label{eq:curlv}
        \end{equation}
        where $\Vec{x}_i = (R_i, \phi_i)$ is the position of each vortex, expressed in polar coordinates in the midplane $z=0$, and $\Vec{e}_z$ is the unit vector along the rotation axis. 
        The integral in equation (\ref{eq:S21}) reduces therefore to a sum over all the vortices in the star, which yields
        \begin{align}
            h_{jk}^{\rm TT} =&\; \left(\frac{512\pi}{405}\right)^{1/2} \frac{G\Omega^2 \rho \kappa R_\ast^4}{Dc^5} T_{jk}^{\rm B2, 2, \pm 1} \nonumber \\
            &\times \sum_{i=1}^{N_{\rm v}} \left(\frac{R_i}{R_\ast}\right)\left(1-\frac{R^2_i}{R_\ast^2}\right)^{3/2} \cos \phi_i ~.
            \label{eq:WaveStrainSum}
        \end{align}
        In equation (\ref{eq:WaveStrainSum}), $T_{jk}^{{\rm B}2, 2, \pm 1}$ are Cartesian components of the tensor spherical harmonic
        \begin{align}
            \textbf{\textsf{T}}^{{\rm B}2, \pm 1} =&\; \left(\frac{5}{32\pi}\right)^{1/2} \bigg[ -\frac{\sin\zeta}{\sin\iota}\left( 1-3\sin^2 \iota\right) \left( \hat{\iota} \otimes \hat{\iota} - \hat{\zeta} \otimes \hat{\zeta} \right) \nonumber \\ 
            &+ \frac{\cos\iota\cos\zeta}{\sin\iota}\left( 1-4\sin^2 \iota\right) \left( \hat{\iota} \otimes \hat{\zeta} + \hat{\zeta} \otimes \hat{\iota} \right) \bigg]~, \label{eq:tensorharmonic}
        \end{align}
        where $\iota$ denotes the inclination angle between the rotation axis of the star and the line of sight, $\zeta$ denotes the azimuth of the line of sight relative to the $\phi=0$ axis, and $\hat{\iota}$ and $\hat{\zeta}$ denote unit vectors in the $\iota$- and $\zeta$-directions, respectively.\footnote{In equations (\ref{eq:Y21})--(\ref{eq:WaveStrainSum}), we use cylindrical polar coordinates $(R, \phi, z)$ motivated by the rectilinearity of the vortices. In equation (\ref{eq:tensorharmonic}), we use spherical polar coordinates $(\iota, \zeta)$, which is the natural choice for specifying the line of sight.}
        Equation (\ref{eq:WaveStrainSum}) corresponds to equation (6) in \citet{Melatosetal2015}. 
        
        The positions of each vortex are known from the N-body simulation, so we can evaluate equation (\ref{eq:WaveStrainSum}) directly. 
        The vortex positions in equation (\ref{eq:WaveStrainSum}) refer to the inertial frame, not the corotating frame, i.e.\ one has $\phi_i = \phi_{i, 0} + \Omega t$, where $\phi_{i, 0}$ is the azimuth of the vortex in the frame corotating with the pinning sites [e.g.\ in Fig. (\ref{fig:GH2020SimDisplay})]. 
        We can therefore write $h_{jk}^{\rm TT} = T_{jk}^{\rm B2, 2, \pm 1}h_0 \cos(\Omega t + \Phi)$ where $\Phi$ is an astrophysically irrelevant phase (which depends on the vortex positions in the corotating frame), with 
        \begin{equation}
            h_0 = \left(\frac{512\pi}{405}\right)^{1/2} \frac{G\Omega^2 \rho \kappa R_\ast^4}{Dc^5} Q ~, 
            \label{eq:WaveStrainFinal}
        \end{equation}
        and
        \begin{align}
            Q =& \left\{ \left[ \sum_{i=1}^{N_{\rm v}} \left(\frac{R_i}{R_\ast}\right)\left(1-\frac{R^2_i}{R_\ast^2}\right)^{3/2}\cos\phi_{i, 0} \right]^2 \right. \nonumber \\ 
             & + \left. \left[ \sum_{i=1}^{N_{\rm v}} \left( \frac{R_i}{R_\ast}\right)\left(1-\frac{R^2_i}{R_\ast^2}\right)^{3/2}\sin\phi_{i, 0} \right]^2 \right\}^{1/2}.
            \label{eq:Q}
        \end{align}

\section{\texorpdfstring{$Q$}{Q} versus \texorpdfstring{$N_{\rm v}$}{Nv}: N-body scaling for \texorpdfstring{$N_{\rm v} \leq 5 \times 10^3$}{Nv leq 5e3}}
    \label{sec:QScalingForSim}
    
    The characteristic wave strain $h_0$ depends on the vortex configuration purely through the dimensionless quantity $Q$ in equations (\ref{eq:WaveStrainFinal}) and (\ref{eq:Q}). 
    The value of $Q$ is constant between avalanches, because the vortices are pinned, but it varies stochastically from one inter-avalanche interval to the next\footnote{It also varies stochastically during every avalanche, as calculated by \citet{WarszawskiMelatos2012} and \citet{DeLilloEtAl2022}, but the latter signal is burst-like (duration $\lesssim 1\,{\rm s}$) and therefore lies outside the scope of this paper, whose focus is continuous gravitational radiation.}. 
    On time-scales that are short compared to the star's spin-down time-scale, the spatial vortex configuration exhibits approximately stationary statistics, and hence so does $Q$. 
    Moreover, Fig. \ref{fig:GH2020SimDisplay} demonstrates that the system does not have a preferred axis, i.e.\ it is isotropic and homogeneous globally albeit not locally. 
    Consequently, the statistics of $Q$, such as its first and second moments $\langle Q \rangle$ and $\langle Q^2 \rangle$ respectively, depend only on $N_{\rm v}$. The prefactor $\propto \Omega^2 \propto N_{\rm v}^2$ in equation (\ref{eq:WaveStrainFinal}) also depends on $N_{\rm v}$, of course, but the latter scaling does not depend on the spatial configuration of the vortices.  
    
    The pinned, metastable vortex configuration between avalanches exemplified by Fig. \ref{fig:GH2020SimDisplay} is a snapshot of a far-from-equilibrium, self-organized critical process. 
    No general analytic theory exists for the statistics of such processes, due to the profound complications introduced by long-range spatiotemporal correlations, ``memory'', and hysteresis \citep{Jensen1998}. 
    In this section, we calculate the scaling of $Q$ versus $N_{\rm v}$ empirically, by evaluating equation (\ref{eq:Q}) from the output of N-body simulations such as the one that produced Fig. \ref{fig:GH2020SimDisplay} \citep{Howittetal2020}. 
    The simulations are restricted to $N_{\rm v} \leq 5 \times 10^3$ due to computational cost. 
    
    \subsection{N-body simulations}
        \label{sec:NbodySim}
        The N-body solver developed by \citet{Howittetal2020} integrates the equations of motion for $N_{\rm v}$ point vortices in two dimensions. 
        The velocity $\dd \Vec{x}_i(t)/\dd t$ of a vortex at $\Vec{x}_i(t)$ equals the bulk velocity of the inviscid superfluid condensate at $\Vec{x}_i(t)$ induced by the other vortices (the N-body term) plus motion arising from interactions with a grid of pinning potentials, the container wall (via image vortices), and a viscous superfluid component [drag and mutual friction, implemented by rotating through a dissipation angle \citep{Schwarz1985}]. 
        The simulations are conducted in the frame that corotates with the stellar crust, which corotates with the pinning grid and viscous component. 
        The angular frequency $\Omega$ of the crust is adjusted at every time step, in response to an external spin-down torque (astrophysically the magnetic dipole braking torque, denoted by $N_{\rm ext}$) and the decrement in the total angular momentum of the superfluid condensate, if one or more vortices change position during the time-step (as happens during an avalanche). 
        The N-body equations of motion and simulation parameters are summarized briefly in Appendix \ref{sec:AppxSimParam} for the reader's convenience.
        
        For the $Q$-versus-$N_{\rm v}$ studies in this section, we run eight independent simulations, each starting with ${N_{\rm v}} = 5\times10^3$. 
        As avalanches occur, and vortices progressively exit the system, we calculate $Q$ across a range of $N_{\rm v}$ values, e.g.\ for $5\times10^2 \leq {N_{\rm v}} \leq 5\times10^3$, as ${N_{\rm v}}$ decreases with time\footnote{Ideally, one would perform this calculation by initialising an excess number of vortices ($1 \times 10^5$, say) at random positions, evolving the system until there are $N_{\rm v}$ vortices left, calculating $Q$ at that value of $N_{\rm v}$, and then repeating for a different initial random seed and $N_{\rm v}$ value, so that every point in the figure would be generated from randomly initialized and independent runs. 
        This is expensive computationally, so we are obliged to partially sacrifice independence and compute $Q$ for multiple $N_{\rm v}$ values during each randomly initialized run. 
        Instead, we combat the interdependence between successive points on a given run by only calculating $Q$ every 10 interglitch intervals, allowing the avalanches that happen in between to ``scramble'' the vortex configuration (see Section 4.2).}. 
        The simulations are initialized by drawing $\{\Vec{x}_i\}$ from a spatially uniform probability distribution. They are evolved with $N_{\rm ext}=0$ until all vortices are pinned, as described in Section 5.1 in \citet{Howittetal2020}. Subsequently, the simulations are evolved with $N_{\rm ext} \neq 0$ in order to trigger avalanches. 
        Avalanches are detected automatically using the glitch-finding algorithm described in Section 3.3 in \citet{Howittetal2020}, after $\Omega(t)$ is smoothed over a few adjacent time-steps with a top-hat function (details about the smoothing are recorded in Appendix \ref{sec:AppxSimParam}). 
        A value of $Q$ representative of each interglitch interval is calculated by evaluating equation (\ref{eq:Q}) using the vortex positions at the midpoint of the interval. 
        
        \begin{figure}
            \centering
            \begin{subfigure}[b]{0.9\linewidth}
                \centering
                \includegraphics[width=\linewidth]{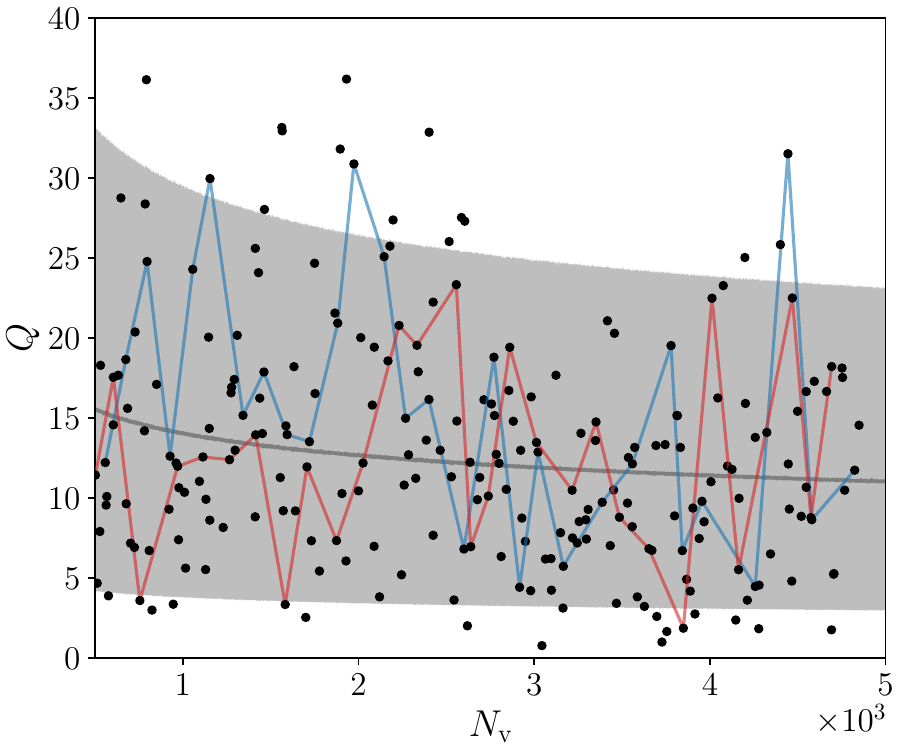}
            \end{subfigure}
            \begin{subfigure}[b]{0.75\columnwidth}
                \centering
                \includegraphics[width=\textwidth]{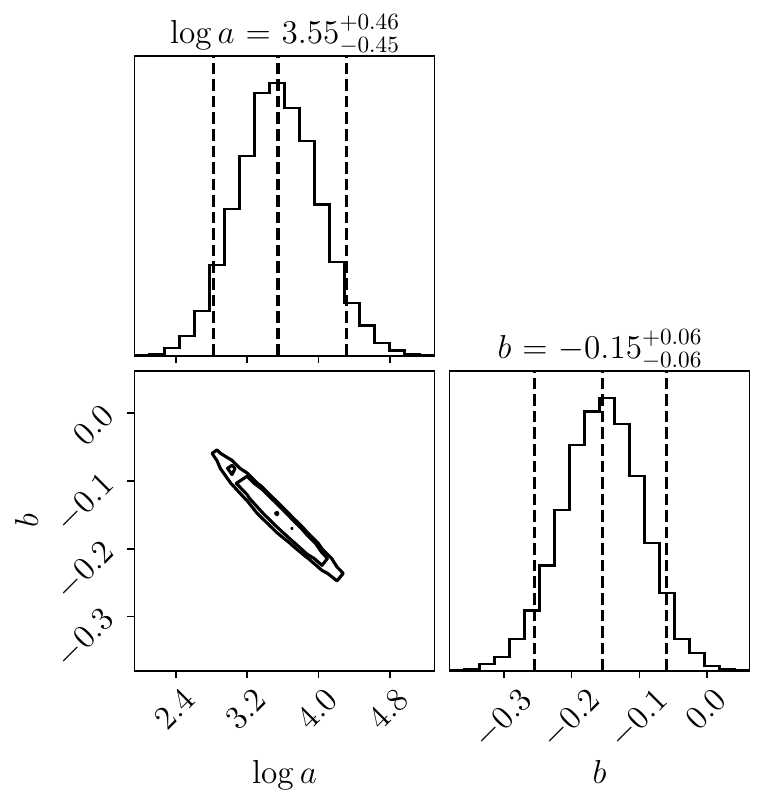}
            \end{subfigure}
            \begin{subfigure}[b]{0.75\columnwidth}
                \centering
                \includegraphics[width=\textwidth]{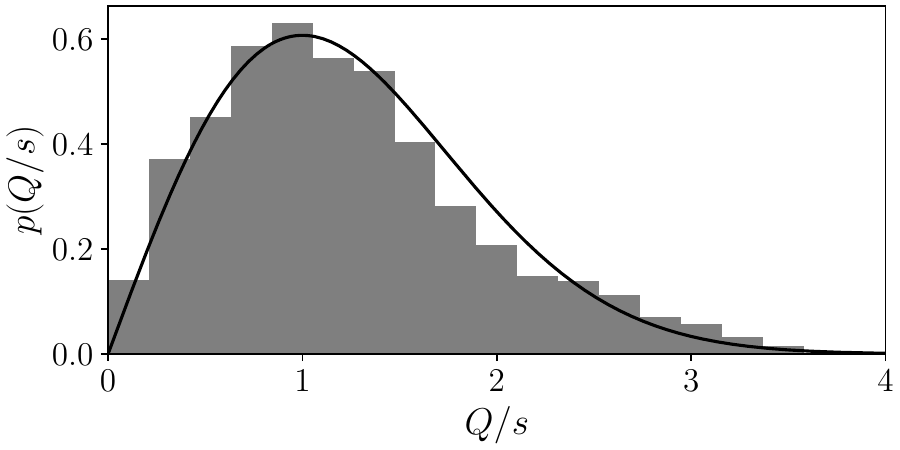}
            \end{subfigure}
            \caption{Statistics of the dimensionless vortex configuration factor $Q$ evaluated in eight independent N-body simulations starting from ${N_{\rm v}} = 5 \times 10^3$ with the parameters in Appendix \ref{sec:AppxSimParam}. 
            \textit{Top panel}: Scatter plot of snapshots $({N_{\rm v}}, Q)$ taken after every 10-th glitch in the range $5 \times 10^2 \leq {N_{\rm v}} \leq 5 \times 10^3$ for all eight runs.  
            $Q$ is calculated by evaluating equation (\ref{eq:Q}) at midpoints between avalanches.
            The red and blue lines each track snapshots from arbitrarily chosen independent runs.
            Snapshots at more than 90 per cent or less than 10 per cent of the initial container frequency are discarded both in the plot and the parameter estimation.
            The dark grey line and the grey band displays the mean and 90 per cent credible interval of the posterior predictive distribution at each $N_{\rm v}$, respectively.             
            \textit{Middle three panels}: Corner plot displaying the two-dimensional and marginalized one-dimensional posterior distributions of the estimated PDF parameters $a$ and $b$ in equation (\ref{eq:Q}). Parameter estimates in the corner plots are reported with their 5-th, 50-th, and 95-th percentiles (vertical dashed lines). 
            \textit{Bottom panel}: Normalized histogram of $Q/s$ with $s=a{N_{\rm v}}^b$ for every $({N_{\rm v}}, Q)$ point in the top panel for $10^3$ samples of $(a, b)$ drawn from the posterior. The Rayleigh distribution $p(Q/s) = (Q/s) \exp{\left[-Q^2/2s^2\right]}$ is overlaid as a solid black curve.}
            \label{fig:Qscaling}
        \end{figure}
    
    \subsection{Probability density function of \texorpdfstring{$Q$}{Q}}
        \label{sec:PDFofQ}
        
        Figure \ref{fig:Qscaling} displays snapshots of $Q$ versus ${N_{\rm v}}$ as a scatter plot. 
        The snapshots are taken at instants within the range $5 \times 10^2 \leq {N_{\rm v}} \leq 5 \times 10^3$ during the eight independent simulations discussed in Section \ref{sec:NbodySim}.
        The instants correspond to the immediate aftermath of every 10-th glitch, starting from the 10-th. 
        That is, the time series $Q(t)$ and ${N_{\rm v}}(t)$ are decimated to ensure that the plotted data points are reasonably independent statistically; 10 avalanches occur between one data point and the next, during which the vortex configuration ``scrambles'' to a degree (see footnote 3).
        Across the eight runs, $Q$ varies by a factor of ${\sim}20$ (smallest to largest) at any fixed ${N_{\rm v}}$ value in the plotted range.
        One may speculate that the initial vortex configuration affects the magnitude of $Q$ for a given run, so that some runs have consistently higher $Q$, and others consistently lower $Q$. 
        We inspect visually the $Q$ of all eight runs, and find that such behaviour is not present, suggesting that $Q$ loses memory of initial conditions. 
        For clarity, we join the dots for only two out of eight runs in Fig. \ref{fig:Qscaling} as the red and blue lines.
        The spread of $Q$ decreases with ${N_{\rm v}}$, as quantified below. 
        There is no strong trend between the central tendency of $Q$ and ${N_{\rm v}}$, but the broad spread in $Q$ makes it hard to identify a trend visually.

        What is the probability density function (PDF) of $Q$ as a function of ${N_{\rm v}}$? We develop an approximate answer to this question by noting from equation (\ref{eq:Q}) that $Q$ is the length of a random vector,
        \begin{equation}
            \Vec{S} = \begin{pmatrix} 
                \sum_{i=1}^{N_{\rm v}} \left(R_i/R_\ast\right)\left(1-R^2_i/R_\ast^2\right)^{3/2}\cos\phi_{i, 0} \\ 
                \sum_{i=1}^{N_{\rm v}} \left(R_i/R_\ast\right)\left(1-R^2_i/R_\ast^2\right)^{3/2}\sin\phi_{i, 0} \end{pmatrix}~, \label{eq:S}
        \end{equation}
        which can be calculated for each vortex configuration. 
        Clearly, each component of $\Vec{S}$ is distributed symmetrically around zero. 
        The exact PDF of each component is challenging to predict analytically, for the reasons discussed in the second paragraph of Section \ref{sec:QScalingForSim}. 
        However, it is unimodal with a steep tail, so we approximate each component of $\Vec{S}$ as a zero-mean Gaussian with standard deviation $s = a {N_{\rm v}}^b$, a power-law parameterization which is deliberately scale-invariant and motivated by the analytically tractable configuration discussed in Section \ref{sec:UniformPoisson}.
        Consequently, at fixed ${N_{\rm v}}$, $Q$ follows a Rayleigh distribution 
        \begin{equation}
            p(Q) = \frac{Q}{a^2 {N_{\rm v}}^{2b}}\exp\left(-\frac{Q^2}{2a^2{N_{\rm v}}^{2b}}\right)~.
            \label{eq:rayleigh}
        \end{equation}

        We sample the posterior distribution of $a$ and $b$ assuming equation (\ref{eq:rayleigh}) and uninformative priors using the Markov Chain Monte Carlo (MCMC) sampler in the \textsc{emcee} Python package \citep{Emcee}. 
        The resulting posterior probability distribution is presented as a corner plot in the middle three panels of Fig. \ref{fig:Qscaling}\footnote{The corner plots are produced by the \textsc{corner.py} Python package \citep{corner}.}.
        We estimate $a = 35\substack{+20 \\ -13}$ and $b = -0.15\substack{+0.06 \\ -0.06}$ where central values correspond to the median, and the error bars delineate the 90 per cent credible intervals of the marginalized posterior distribution.
        
        The dark grey line and the grey band in the top panel of Fig. \ref{fig:Qscaling} indicate the median and the 90 per cent credible interval of the posterior predictive distribution at each $N_{\rm v}$, respectively. 
        This posterior predictive distribution includes both the scatter due to avalanche stochasticity and uncertainty in parameter estimates. 

        A natural question is to ask whether the model in equation (\ref{eq:rayleigh}) describes the data well. 
        Ideally, we would perform a large number (${\gtrsim}10^3$, say) of simulations at each $N_{\rm v}$, to compare an empirical distribution of $Q$ at a given $N_{\rm v}$ with equation (\ref{eq:rayleigh}) directly. 
        However, this is computationally prohibitive. 
        Instead, in the bottom panel of Fig. \ref{fig:Qscaling}, we perform the following quantitative consistency check, using data at all values of $N_{\rm v}$ simultaneously. 
        We sample $10^3$ values of $(a, b)$ from the posterior probability distribution. 
        For each sample, and each datum in the top panel of Fig. \ref{fig:Qscaling}, we compute $Q / s$, with $s = a N_{\rm v}^b$. 
        We compare a histogram of all such values with a Rayleigh distribution described by $p(Q / s) = (Q / s)\exp \left[-Q^2 /(2s^2) \right]$ in the bottom panel of Fig. \ref{fig:Qscaling} (grey bars and solid black curve respectively). 
        If the distribution of $Q$ at each $N_{\rm v}$ deviates significantly from equation (\ref{eq:rayleigh}), the empirical histogram of $Q / s$ would fail to follow the Rayleigh curve. 
        By visual inspection, the histogram and theoretical distribution are broadly consistent, giving us some confidence that equation (\ref{eq:rayleigh}) describes the data well.

        In an ensemble of vortex configurations described by equation (\ref{eq:rayleigh}) with median estimates of $a$ and $b$, a typical $Q$ is given by
        \begin{equation}
            Q = 41\substack{+44 \\ -30} ~ {N_{\rm v}}^{-0.15} \label{eq:Nbodyscaling}
        \end{equation}
        where the central value and error bars correspond to the median and the central 90 percentile of equation (\ref{eq:rayleigh}), respectively. 
        It is important to note that a posterior distribution is not a point estimate, but we derive equation (\ref{eq:Nbodyscaling}) using the median of the posterior distribution to quantify the scatter due to stochasticity of the avalanches. 
        This scatter is distinct from the credible interval for the estimates of $a$ and $b$ (vertical dashed lines in the middle three panels of Fig. \ref{fig:Qscaling}).
        At a given ${N_{\rm v}}$, there are many possible configurations of vortices depending on the random initial configuration and avalanche history.
        Individual configurations typically fall within the error bars of equation (\ref{eq:Nbodyscaling}), and there is no way to know where the actual configuration for a specific astronomical object lies within the range.
        We emphasize that equation (\ref{eq:Nbodyscaling}) is not the posterior predictive distribution displayed as the grey band in the top panel of Figure \ref{fig:Qscaling}. 
        Rather, it can be thought of as a best-fit distribution of equation (\ref{eq:rayleigh}) to the data, which is quoted in order to distinguish the uncertainty in the estimates of $a$ and $b$ from the inherent stochasticity and hence dispersion of the avalanches which generate the vortex configurations.

\section{\texorpdfstring{$Q$}{Q} versus \texorpdfstring{$N_{\rm v}$}{Nv}: Extrapolating to realistic \texorpdfstring{$N_{\rm v} \gg 10^4$}{Nv >> 1e4}}
    \label{sec:QScalingForConfigs}

    The N-body simulations in Section \ref{sec:QScalingForSim} are restricted to ${N_{\rm v}} \leq 5\times10^3$ by computational cost, whereas a neutron star contains ${\sim}10^{15}$ ($\Omega /1\, {\rm rad\, s^{-1}}$) vortices. The $Q$-versus-${N_{\rm v}}$ data in Fig. \ref{fig:Qscaling} are approximated well by equations (\ref{eq:rayleigh}) and (\ref{eq:Nbodyscaling}) over one decade in ${N_{\rm v}}$. 
    This empirical fact offers some encouragement.
    Needless to say, extrapolating over 12 decades of ${N_{\rm v}}$ into the neutron star regime is dangerous even for scale-invariant, power-law expressions like equation (\ref{eq:Nbodyscaling}). 
    One would prefer to have a theoretical justification for the empirical scaling (\ref{eq:Nbodyscaling}), but such a justification is hard to derive from first principles, due to the well-documented and unsolved theoretical challenges posed by the statistical mechanics of far-from-equilibrium systems, as discussed in Section \ref{sec:SuperfluidVortexAvalanches} \citep{Jensen1998}. 
    Instead, in this section, we calculate $Q$ as a function of ${N_{\rm v}}$ for two analytically tractable point processes which ``bracket'' the self-organized critical dynamics observed in the simulated vortex avalanches in Section \ref{sec:QScalingForSim}, i.e.\ which bound the simulated $Q$ above and below.

    The uniform Poisson point process (Section \ref{sec:UniformPoisson}), where vortex positions are drawn independently from a uniform probability distribution, approximates the regime where vortices do not repel each other, and vortex positions are uncorrelated. 
    The uniform Poisson configuration is less symmetric than the avalanche configuration in Section \ref{sec:QScalingForSim} and produces median $Q$ greater than the central value in equation (\ref{eq:Nbodyscaling}). 
    The regular array configuration (Section \ref{sec:RegularArray}) realizes the other extreme, where vortex repulsion dominates pinning, and vortex positions are highly correlated; in fact, the vortices are arranged periodically in an Abrikosov-like array. 
    The regular array configuration is more symmetric than the avalanche configuration in Section \ref{sec:QScalingForSim} and produces median $Q$ less than the central value in equation (\ref{eq:Nbodyscaling}). 
    The analytic $Q$-versus-${N_{\rm v}}$ scalings in Sections \ref{sec:UniformPoisson} and \ref{sec:RegularArray} bracket the avalanche-driven N-body scaling and can therefore be employed to place bounds on $h_0$ in astrophysical applications, where conservatism is preferred, and the extrapolation of equation (\ref{eq:Nbodyscaling}) is deemed unreliable\footnote{The least symmetric vortex configuration with the highest $Q$ corresponds to placing every vortex at one off-axis point at radius $R$. This artificial configuration has $Q=(R/R_\ast)(1 - R^2/R_\ast^2)^{3/2}{N_{\rm v}}$ from equation (\ref{eq:Q}) but is not relevant to a neutron star.}.

    \subsection{Uniform Poisson configuration: upper bound on \texorpdfstring{$Q$}{Q}}
        \label{sec:UniformPoisson}
        A uniform Poisson point process generates a configuration of points in a two-dimensional region by placing each point at an independent location, with all locations having equal probability per unit area. 
        In this paper, where vortices are assumed to be rectilinear, the region of interest is an equatorial disk of radius $R_{\ast}$. 
        Of course, this is not a realistic generative model for vortex configurations such as in Fig. \ref{fig:GH2020SimDisplay}. 
        Vortex positions resulting from avalanches are correlated due to vortex-vortex repulsion and knock-on unpinning, whereas the uniform Poisson point process involves zero vortex-vortex repulsion and zero correlation between vortex positions. 
        For these reasons, a random realization of a uniform Poisson point process is less symmetric typically, with higher $Q$, than a realistic avalanche snapshot like Fig. \ref{fig:GH2020SimDisplay}. 
        One can calculate the PDF and moments of $Q$ analytically for arbitrary ${N_{\rm v}}$ and derive an upper bound on $Q$ for astrophysical applications.
        
        Consider the random vector $\Vec{S}$ defined in equation (\ref{eq:S}). We show in Appendix \ref{sec:AppxPoisson} using the central limit theorem that the uniform Poisson point process leads to each component of $\Vec{S}$ being distributed according to a normal distribution with standard deviation $({N_{\rm v}}/40)^{1/2}$, viz.
        \begin{equation}
            p(S_x) = \left(\frac{20}{\pi {N_{\rm v}}}\right)^{1/2} \exp{\left(-\frac{20S_x^2}{{N_{\rm v}}}\right)}
        \end{equation}
        and similarly for $p(S_y)$, where $x$ and $y$ denote Cartesian coordinates. 
        Therefore, the PDF for $Q = |\Vec{S}|$ is the Rayleigh distribution 
        \begin{equation}
            p(Q) = \left( \frac{40Q^2}{{N_{\rm v}}} \right)^{1/2}  \exp{\left(-\frac{20 Q^2}{{N_{\rm v}}}\right)} , \label{eq:Q_rayleigh_uniform}   
        \end{equation}
        which matches equation (\ref{eq:rayleigh}) with $a=(1/40)^{1/2}$ and $b=1/2$. 
        The median and central 90 percentile are given by
        \begin{equation}
            Q = 0.18 \substack{+0.20 \\ -0.13} ~ {N_{\rm v}}^{0.50} ~. \label{eq:Uniformscaling}
        \end{equation}

        \begin{figure}
            \centering
            \includegraphics[width=\linewidth]{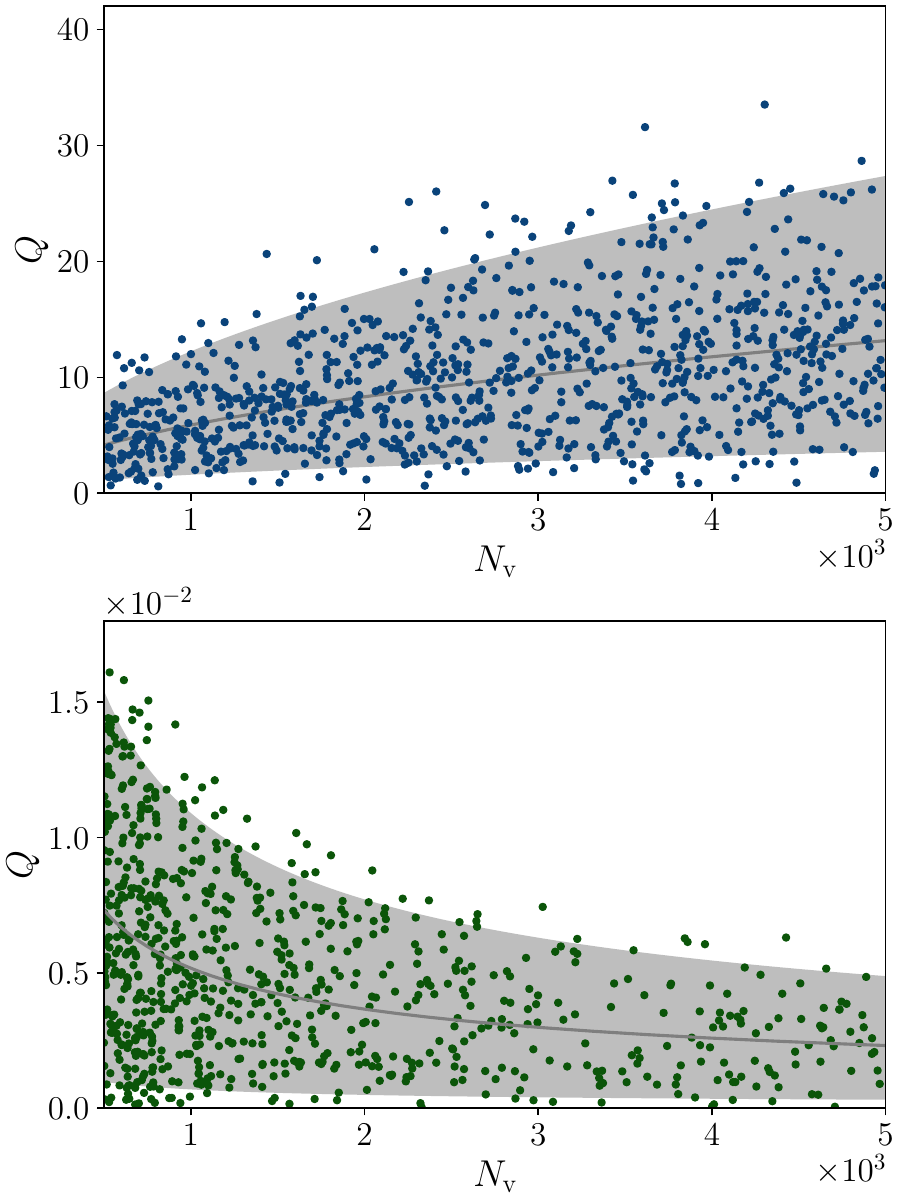}
            \caption{Scaling of the dimensionless vortex configuration factor $Q$ as a function of the number of vortices $N_{\rm v}$. \textit{Top panel}: Samples of $Q$ for $10^3$ random configurations generated by a uniform Poisson point process for $5\times10^2 \leq {N_{\rm v}} \leq 5\times10^3$. For each configuration, $N_{\rm v}$ is drawn from a discrete uniform distribution on the interval $[5\times10^2, 5\times10^3]$, and $Q$ is calculated by evaluating equation (\ref{eq:Q}) for each configuration. The dark grey curve correspond to the median of the theoretical PDF in equation (\ref{eq:Q_rayleigh_uniform}) at each $N_{\rm v}$. The grey band shows the central 90 percentile of the same PDF. \textit{Bottom panel}: Samples of $Q$ for $10^3$ realizations of a regular periodic array with randomized centre for $5\times10^2 \leq {N_{\rm v}} \leq 5\times10^3$. For each configuration, the intervortex spacing $d$ is drawn from a continuous uniform distribution on the interval $[2\times10^{-2}, 9\times10^{-2}]$, $c_x$ and $c_y$ are drawn from a uniform distribution on $[0, d)$, and $Q$ is calculated by evaluating equation (\ref{eq:Q}) for each configuration.}
            \label{fig:uniform_and_array}
        \end{figure}
        
        Let us verify the above scaling numerically. 
        The top panel of Fig. \ref{fig:uniform_and_array} shows the $Q$-versus-${N_{\rm v}}$ scaling for a randomly realized set of vortex configurations generated by the uniform Poisson point process. 
        Configurations are generated for $10^3$ samples of $N_{\rm v}$ drawn from a discrete uniform distribution on the interval $[5\times10^2, 5\times10^3]$, with $Q$ calculated via equation (\ref{eq:Q}). 
        The dark grey line and grey band in the top panel of Fig. \ref{fig:uniform_and_array} displays the median and the central 90 percentile of equation (\ref{eq:Q_rayleigh_uniform}), respectively.
        Approximately 90 per cent of the data lie within the grey band.
        As a quantitative consistency check, we use an MCMC sampler in the \textsc{emcee} Python package to sample the posterior distribution of $a$ and $b$ assuming equation (\ref{eq:rayleigh}) and uninformative priors. 
        We find the 90 per cent credible intervals of the marginalized posterior distribution of $a$ and $b$ to be $[0.13, 0.25]$ and $[-0.53, -0.44]$ respectively.
        These credible intervals contain the theoretical values $a\approx0.15$ and $b = 0.5$, validating equation (\ref{eq:Q_rayleigh_uniform}).
    
    \subsection{Regular array: lower bound on \texorpdfstring{$Q$}{Q}}
        \label{sec:RegularArray}
        The uniform Poisson point process produces zero correlation between vortex positions. 
        It cannot be responsible for spatially correlated configurations containing filaments and voids, such as the one in Fig. \ref{fig:GH2020SimDisplay}. 
        Long-range spatial correlations are a standard feature in self-organized critical systems such as sand piles \citep{Jensen1998}. In the application studied here, they arise from vortex-vortex repulsion and the randomizing action of avalanches, which lead to capacitive depletion zones as discussed by previous authors \citep{ChengEtAl1988, AlparEtAl1996, AlparBaykal2006, MelatosWarszawski2009}. 
        The uniform Poisson process in Section \ref{sec:UniformPoisson} leads to a higher value of $Q$ than for spatially correlated configurations, such as the one in Fig. \ref{fig:GH2020SimDisplay}. 
        
        One can place a lower bound on $Q$ for spatially correlated vortex positions by considering regular configurations dominated by vortex-vortex repulsion. 
        In the latter regime, the vortices form a periodic array akin to an Abrikosov vortex array.  
        The periodicity causes near-cancellation between vortices reflected about the origin. 
        The cancellation becomes more perfect as ${N_{\rm v}}$ increases. 
        
        In Appendix \ref{sec:AppxArray}, we derive an exact, closed-form expression for $Q$ as a function of ${N_{\rm v}}$ for a regular square array with intervortex spacing $d$ and centre $(c_x, c_y)$ relative to the origin. 
        The square array is not necessarily the same as the true equilibrium configuration in the complicated environment of the neutron star, but it is a reasonable and analytically tractable approximation. 
        We note that the vortex array need not match the symmetry of pinning sites in general, as demonstrated by the vortex configurations in Section \ref{sec:QScalingForSim} and \citet{Howittetal2020}. 
        In Appendix \ref{sec:AppxArray}, we find $Q=|\tilde{S}|$ with
        \begin{align}
            \tilde{S} =\,& \frac{3}{2\pi} \sum_{m_1 = -\infty}^{\infty} \sum_{m_2 = -\infty}^{\infty} \frac{j_3\left[2 \pi R_\ast (m_1^2 + m_2^2)^{1/2}/d\right]}{m_1^2 + m_2^2} \nonumber \\
            &\times \exp{\left\{ {\rm i} \left[\frac{2\pi}{d}(m_1 c_x + m_2 c_y) + \tan^{-1}\left(\frac{m_2}{m_1}\right)\right]\right\}} ~, \label{eq:Stilde_array}
        \end{align}
        where $j_n(z)$ is the spherical Bessel function of order $n$. 
        Unlike the uniform Poisson configuration of Section \ref{sec:UniformPoisson}, a regular array configuration is not stochastic; given $(c_x, c_y)$, $Q$ can be computed deterministically from equation (\ref{eq:Stilde_array}). 
        
        It is clear from equation (\ref{eq:Stilde_array}) that $Q$ vanishes for special choices of the centre, such as $(c_x, c_y)=(0, 0)$. 
        This follows intuitively from equation (\ref{eq:Q}). 
        If the regular array is centred at the origin, every vortex at $(R_i, \phi_{i, 0})$ has a counterpart at $(R_i, \phi_i + \pi)$. 
        The terms corresponding to these two vortices cancel perfectly in the sums in equation (\ref{eq:Q}). 
        Such perfect cancellation relies on the exact periodicity of the configuration, and would be unlikely to occur in a neutron star. 
        In general, one has $0 < c_j < d$, where $j \in \{x, y\}$, the cancellation is imperfect, and one finds $Q \neq 0$. 
        We show in Appendix \ref{sec:AppxArray} that one obtains $Q \propto {N_{\rm v}}^{-1/2}$ for ${N_{\rm v}} \gg 1$. Notably, the scaling applies to a regular array of any shape, not just a square array.
  
        Given $R_\ast$ and $d$, $Q$ ranges from zero to a maximum value depending on $(c_x, c_y)$. 
        In this paper, we explore the range of $Q$ by drawing $c_x$ and $c_y$ from a uniform distribution on the interval $[0, d)$. 
        The bottom panel of Fig. \ref{fig:uniform_and_array} shows the $Q$-versus-${N_{\rm v}}$ scaling for a set of regular array configurations, with randomized $(c_x, c_y)$. 
        Vortex configurations are generated for a set of $10^3$ samples of $d$ drawn from a continuous uniform distribution on the interval $[2\times10^{-2}, 9\times10^{-2}]$.
        For each vortex configuration, ${N_{\rm v}}$ is counted, and $Q$ is computed using equation (\ref{eq:Q}). 
        To match the range of ${N_{\rm v}}$ in Sections \ref{sec:QScalingForSim} and \ref{sec:UniformPoisson}, only vortex configurations with $5\times10^2 \leq {N_{\rm v}} \leq 5\times10^3$ are included in the plot.

        Since the scatter of $Q$ in the bottom panel of Fig. \ref{fig:uniform_and_array} is a result of randomizing $(c_x, c_y)$, not the stochasticity of the vortex configurations, the distribution of $Q$ at each $N_{\rm v}$ does not follow equation (\ref{eq:rayleigh}). 
        Deriving the analytic PDF that characterizes this scatter is difficult, due to the complicated form of equation (\ref{eq:Stilde_array}). 
        Instead, we calculate numerically a set of heuristic scalings comparable to the typical $Q$ of equations (\ref{eq:Nbodyscaling}) and (\ref{eq:Uniformscaling}). 
        Motivated by the analysis in Appendix \ref{sec:AppxArray}, we consider a family of curves $Q = a{N_{\rm v}}^{-1/2}$, and find numerically the values of $a$ that lead to curves which lie above 5 per cent, 50 per cent, and 95 per cent of the data in the bottom panel of Fig. \ref{fig:uniform_and_array}. 
        We find empirically that the typical $Q$ for a regular array configuration is
        \begin{equation}
            Q = 0.16\substack{+0.18 \\ -0.14} ~ N_{\rm v}^{-0.5} \label{eq:Arrayscaling}
        \end{equation}
        where, by construction, the central value bisects the data, and the error bars bracket the central 90 percentile.
        We plot the central value and the uncertainty in equation (\ref{eq:Arrayscaling}) in Fig. \ref{fig:uniform_and_array} as the dark grey line, and the grey band respectively.

    \subsection{Pinning potential}
        \label{sec:PinningPotential}
    
        The $Q$-versus-${N_{\rm v}}$ bounds in Sections \ref{sec:UniformPoisson} and \ref{sec:RegularArray} are calculated via geometric arguments and therefore do not depend on the pinning potential $V_0$ (see equation (\ref{eq:pinningpotential}) and Table \ref{tab:SimParams}), as long as $V_0$ occupies the astrophysically relevant regime, where vortex avalanches (and hence rotational glitches) occur. 
        If $V_0$ is too low, then vortices are weakly pinned. As the crust decelerates, the Magnus force pulls vortices outward one-by-one, and few vortices in the configuration are on the verge of unpinning at any instant. 
        Hence, vortices unpinned by knock-on move slowly and repin before they knock-on other vortices, and avalanches are curtailed. 
        Angular momentum transfers gradually to the crust, the superfluid closely matches the angular velocity of the crust at all times, and there is no glitch. 
        If $V_0$ is too high, then vortices cannot unpin until the angular velocity differential becomes large, and one large avalanche occurs\footnote{In practice, the computation takes too long ($\gtrsim$ 1 month) for $V_0 \gtrsim 20$ in normalized code units.}.
        
        It is interesting to ask how the value of $V_0$ in normalized code units compares with the values of $V_0$ in unnormalized, physical units in a realistic neutron star. 
        It is clear from the outset that one should not expect agreement, because the length- and time-scales in the simulation are far from the neutron star regime for computational tractability. 
        Specifically, we have (\citealp{LinkEpstein1991}, see also equation (43) in \citealp{AntonelliHaskell2020})
        \begin{equation}
            V_0  = \frac{E_{\rm p}}{\rho \kappa a_{\rm pin}}~, \label{eq:V0scaling}
        \end{equation}
        where $E_{\rm p}$ is the interaction energy per nucleus, and $a_{\rm pin}$ is the pinning lattice spacing. 
        The simulations in Section \ref{sec:QScalingForSim} have $V_0=2$ in normalized code units, which corresponds to $V_0 = \kappa/\pi$ in physical units and hence
        \begin{equation}
            E_{\rm p} = 2.4\times10^2~{\rm MeV} \left(\frac{a_{\rm pin}}{30 ~{\rm fm}}\right) \label{eq:Ep}
        \end{equation}
        This is two orders of magnitude larger than the typical predictions ($E_{\rm p} \sim$ few MeV) of quantum simulations \citep{LinkEpstein1991, DonatiPizzochero2006, WlazlowskiEtAl2016}, and the values inferred from glitch activity \citep{GugercinogluEtAl2022, MelatosMillhouse2023}.
        This confirms that the simulations are outside the neutron star regime, as expected. 
        Note that $V_0=2$ is close to the minimum pinning potential where avalanches still occur in the N-body solver \citep{Howittetal2020}.

        \citet{SevesoEtAl2016} argued that the effective pinning strength for long vortex lines (${\sim}10^3 a_{\rm pin}$) is less than typical estimates due to the varying orientation of the pinning lattice along the vortex line.
        They proposed the alternative scaling
        \begin{equation}
            V_0 = \frac{\alpha E_{\rm p}}{\rho \kappa a_{\rm pin}}~, \label{eq:newV0scaling}
        \end{equation}
        i.e.\ equation (44) in \citet{AntonelliHaskell2020}, where $10^{-3} \lesssim \alpha \lesssim 10^{-1}$ is a dimensionless factor quantifying the reduction in pinning force.
        Upon replacing equation (\ref{eq:V0scaling}) with equation (\ref{eq:newV0scaling}), we find that $E_{\rm p}$ in equation (\ref{eq:Ep}) is increased by one to three orders-of-magnitude. 
        Therefore, if the arguments of \citet{SevesoEtAl2016} apply astrophysically, then the simulations move further from the neutron star regime.

\section{Characteristic Wave Strain}
    \label{sec:astrowavestrain}
    
    There is sufficient uncertainty about vortex microphysics in neutron stars that it is worthwhile to evaluate the gravitational wave amplitude for all three $Q$-versus-${N_{\rm v}}$ scalings in Sections \ref{sec:QScalingForSim} and \ref{sec:QScalingForConfigs}, that is, equation (\ref{eq:Nbodyscaling}) for extrapolated numerical simulations of vortex avalanches, equation (\ref{eq:Uniformscaling}) for a uniform Poisson configuration, and equation (\ref{eq:Arrayscaling}) for a regular array.
    It is likely that the avalanche $h_0$ lies between the uniform Poisson and regular array results. 
    The wide range is a fair reflection of the uncertainties surrounding the subtle far-from-equilibrium physics of vortex avalanches \citep{Jensen1998}.
    
    Let us begin by considering the extrapolated simulation scaling. 
    Upon substituting $Q$ from equation (\ref{eq:Nbodyscaling}) into equation (\ref{eq:WaveStrainFinal}) and applying the Feynman condition $\kappa {N_{\rm v}} = 2\pi R_\ast^2 \Omega = 4 \pi^2 R_\ast^2 f$ \citep{AndersonItoh1975}, we obtain
    \begin{align}
        h_0 =\;& 7.3 \substack{+7.9 \\ -5.4}\, \times 10^{-42} \left(\frac{f}{30 \;{\rm Hz}}\right)^{1.9} \nonumber\\ 
        &\times\left(\frac{R_\ast}{10 \;{\rm km}}\right)^{0.7} \left(\frac{M}{1.4 M_{\odot}}\right)\left(\frac{D}{1 \;{\rm kpc}}\right)^{-1} ~. \label{eq:SimStrainEstimate}
    \end{align}
    The central value in equation (\ref{eq:SimStrainEstimate}) refers to the median $h_0$ over random realizations of the simulated vortex configuration, and the error bars correspond to the central 90 percentile.
    In an astronomical context, one observes a single realization (namely the actual one), so the median is not observable directly. 
    Instead, the characteristic wave strain for a given interglitch interval is sampled from a Rayleigh distribution with median given by the central value of equation (\ref{eq:SimStrainEstimate}). 
    Equation (\ref{eq:SimStrainEstimate}) is ${\sim}10$ orders of magnitude lower than the central-limit-theorem estimate calculated in Section 5 of \citet{Melatosetal2015}, which assumes $\langle h_0 \rangle \propto \Omega^2 N_{\rm v}^{-1/2}$. 
    
    The upper bound corresponding to a uniform Poisson configuration (Section \ref{sec:UniformPoisson}) is calculated by substituting $Q$ from equation (\ref{eq:Uniformscaling}) into equation (\ref{eq:WaveStrainFinal}) to obtain
    \begin{align}
        h_0 =\;& 1.2 \substack{+1.3 \\ -0.9}\, \times 10^{-32} \left(\frac{f}{30 \;{\rm Hz}}\right)^{2.5} \nonumber\\ 
        &\times\left(\frac{R_\ast}{10 \;{\rm km}}\right)^{2.0} \left(\frac{M}{1.4 M_{\odot}}\right) \left(\frac{D}{1 \;{\rm kpc}}\right)^{-1} ~. \label{eq:UniformStrainEstimate}
    \end{align}
    The lower bound corresponding to a regular array configuration (Section \ref{sec:RegularArray}) is calculated by substituting $Q$ from equation (\ref{eq:Arrayscaling}) into equation (\ref{eq:WaveStrainFinal}), to obtain
    \begin{align}
        h_0 =\;& 1.8 \substack{+2.0 \\ -1.5}\, \times 10^{-50} \left(\frac{f}{30 \;{\rm Hz}}\right)^{1.5} \left(\frac{M}{1.4 M_{\odot}}\right) \left(\frac{D}{1 \;{\rm kpc}}\right)^{-1} ~. \label{eq:ArrayStrainEstimate}
    \end{align}

\section{Conclusions}
    Neutron star glitches may be caused by the sudden unpinning and collective movement of vortices in the superfluid condensate inside the star, also known as vortex avalanches. 
    The metastable vortex configuration between avalanches is determined by the far-from-equilibrium avalanche dynamics and is nonaxisymmetric in general, producing a small but nonzero current quadrupole moment which generates gravitational waves as it rotates \citep{Melatosetal2015}. 
    In this paper, we use the N-body solver developed by \citet{Howittetal2020} to simulate vortex avalanches with $N_{\rm v} \lesssim 5\times10^3$ vortices. 
    We find that the current quadrupole moment scales $\propto {N_{\rm v}}^{-0.15}$, implying $h_0 = 7.3\substack{+7.9 \\ -5.4} \times 10^{-42} (M/1.4 M_\odot)(f/30\,{\rm Hz})^{1.9}(R_\ast/10\,{\rm km})^{0.7}(D/1\,{\rm kpc})^{-1}$ upon extrapolating to the neutron star regime, with ${N_{\rm v}} \sim 10^{17} (f / 30\,{\rm Hz})$.  
    Cautious about extrapolating over ${\gtrsim}10$ decades in $N_{\rm v}$, we also develop for safety two analytic scalings for the uniform Poisson and the regular array configurations, which represent upper and lower bounds on $h_0$ given by equations (\ref{eq:UniformStrainEstimate}) and (\ref{eq:ArrayStrainEstimate}) respectively. 
    The bounds correspond to vortex pinning dominating mutual repulsion (uniform Poisson) and vice versa (regular array) and bracket the empirical scaling extrapolated from the simulations. 
    The wide range is a fair reflection of the uncertainties surrounding the subtle question of long-range correlations between vortex positions in self-organized critical systems like vortex avalanches, which remains a fundamental unsolved problem in statistical mechanics. 
    Due to the stochasticity of vortex avalanches, the current quadrupole moment at a fixed $N_{\rm v}$ scatters around an average according to a distribution that is well-approximated by the Rayleigh distribution [equation (\ref{eq:rayleigh})].
    
    The mechanism in this paper predicts the wave strain to increase strongly with pulsar frequency. 
    Consequently, millisecond pulsars, which have $f \gtrsim 100$ Hz, are favored as search candidates.
    For a millisecond pulsar with $f=500 \, {\rm Hz}$, equation (\ref{eq:SimStrainEstimate}) predicts median wave strain $h_0 = 1.3 \times 10^{-39}$. 
    If vortex dynamics are dominated by pinning, then the appropriate scaling is equation (\ref{eq:UniformStrainEstimate}), which predicts median wave strain $h_0 = 1.4 \times 10^{-29}$.
    Recent narrow-band searches using data from the LIGO-Virgo third observing run place an upper limit for continuous gravitational waves from known pulsars at $h_0 \lesssim 2\times10^{-26}$ \citep{CWsearch}.
    However, it is unclear if vortex avalanches occur in most millisecond pulsars. 
    Only two glitches have ever been observed in millisecond pulsars, one in PSR J1824–2452 \citep{CognardBacker2004} and the other in PSR J0613-0200 \citep{McKeeEtAl2016}. 
    If the paucity of glitches in millisecond pulsars is because some unknown physics intervenes to prevent vortex avalanches (e.g.\ due to a temperature or age threshold excluded from the N-body simulations in Section \ref{sec:NbodySim}), then the prospects for detection diminish accordingly. 
    On the other hand, it may be that most millisecond pulsars do glitch, but only a few glitches have been detected because the average waiting time between events is longer than in ordinary pulsars, e.g.\ once every $10^3$ years instead of once per year \citep{ShemarLyne1996, LyneEtAl2000, EspinozaEtAl2011, MillhouseEtAl2022}. 
    In the latter scenario, the avalanche process is ongoing (albeit slowly), and millisecond pulsars may have nonaxisymmetric vortex distributions like in Fig. \ref{fig:GH2020SimDisplay} today, while they are being observed. 
    This scenario is explored in Section 4.2 of \citet{Melatosetal2015}. 
    If it is viable, millisecond pulsars are plausible targets for continuous gravitational wave searches \citep{AbbottEtAl2010, VargasMelatos2022, AbbottEtAl2022A, AbbottEtAl2022B}.

    The two-dimensional simulations and analytic calculations in this paper assume that vortices are rectilinear. 
    In reality, the vortices in a neutron star are likely to be curved, whereupon vortex tension plays a vital role \citep{LinkEpstein1991, HirasawaShibazaki2001}.    
    Furthermore, the vortices in a neutron star are likely to be tangled, both because differential rotation drives macroscopic, Kolmogorov-like turbulence \citep{Greenstein1970, PeraltaEtAl2005, Peraltaetal2006, PeraltaMelatos2009, MelatosPeralta2010, KhomenkoEtAl2019}, and because pinning and differential rotation combine to generate a microscopic vortex tangle via Kelvin-wave and other instabilities \citep{GlabersonJohnsonOstermeier1974, Donnelly1991, AnderssonEtAl2007, MongioviEtAl2017, DrummondMelatos2017, DrummondMelatos2018, HaskellEtAl2020, ThongEtAl2023, LevinLink2023}. 
    It is unclear whether turbulence increases or decreases $h_0$. 
    On the one hand, it may be argued that the randomizing and hence homogenizing action of eddy-like motions overwhelms mutual vortex repulsion and pushes the system towards a uniform Poisson configuration, increasing $h_0$ relative to equation (\ref{eq:SimStrainEstimate}). 
    On the other hand, it may be argued that eddy-like motions imprint macroscopic, eddy-related length-scales on the vortex configuration, which are unrelated to and much larger than the microscopic vortex separation. 
    The eddies are axisymmetric when averaged over time but not instantaneously, so they contribute to the current quadrupole moment and indeed introduce another time-scale, the eddy turnover time-scale, which competes with $f^{-1}$. 
    Previous studies of gravitational waves from neutron star turbulence are purely hydrodynamic \citep{MelatosPeralta2010, LaskyEtAl2013}; they analyze the eddies only, not the quantized vortices that the eddies contain. 
    The combined problem is subtle, requires more sophisticated and expensive simulations to be studied with confidence, and lies outside the scope of this paper.

\section*{Acknowledgements}
The authors thank Kok Hong Thong for helpful discussions. We acknowledge support from the Australian Research Council (ARC) through the Centre of Excellence for Gravitational Wave Discovery (OzGrav) (grant number CE170100004), an ARC Discovery Project (grant number DP170103625), and the Haasz Family Fund. This research was supported by The University of Melbourne’s Research Computing Services and the Petascale Campus Initiative. 

\section*{Data Availability}

Data used in this work can be made available on reasonable request to the corresponding author.



\bibliographystyle{mnras}
\bibliography{bibliography} 



\appendix

\section{Simulation parameters and data analysis methods}
    \label{sec:AppxSimParam}
    For completeness, we include the equations of motion for the N-body simulation, and a short explanation of relevant parameters. A detailed description can be found in sections 2 and 3 of \citet{Howittetal2020}. 
    
    Let us consider rectilinear vortices in an infinitely long cylindrical container.
    Let the symmetry axis of the container be the $z$-axis. 
    In the frame of reference corotating with the container, we denote the position of the vortex $i$ by $(x_i, y_i)$. 
    The equations of motion of vortex $i$ are
    \begin{equation}
        \frac{\dd}{\dd t} \begin{pmatrix}
            x_i \\ y_i
        \end{pmatrix} 
        = \mathcal{R}_{\phi} \begin{pmatrix}
            v_{i,x} \\ v_{i,y}
        \end{pmatrix} 
        \label{eq:EOMall}
    \end{equation}
    with 
    \begin{align}
        v_{i,x} &= - \sum_{\substack{j = 1 \\ j \neq i}}^{N_{\rm v}} \frac{\kappa y_{ij}}{r_{ij}^2} + \sum_{j = 1}^{{N_{\rm v}}} \frac{\kappa y_{ij, {\rm im}}}{r_{ij, {\rm im}}^2} + \Omega y_i - \sum_{k=1}^{N_{\rm pin}} \frac{\partial V(\Vec{x}_i - \Vec{x}_k)}{\partial y_i} \label{eq:EOMx}\\
        v_{i,y} &= \sum_{\substack{j = 1 \\ j \neq i}}^{N_{\rm v}} \frac{\kappa x_{ij}}{r_{ij}^2} - \sum_{j = 1}^{{N_{\rm v}}} \frac{\kappa x_{ij, {\rm im}}}{r_{ij, {\rm im}}^2} - \Omega x_i + \sum_{k=1}^{N_{\rm pin}} \frac{\partial V(\Vec{x}_i - \Vec{x}_k)}{\partial x_i}~. \label{eq:EOMy}
    \end{align}
    The first terms of equations (\ref{eq:EOMx}) and (\ref{eq:EOMy}) are the induced velocities at vortex $i$ due to other vortices, with $\Vec{x}_{ij} = \Vec{x}_{i} - \Vec{x}_{j} = (x_{ij}, y_{ij})$ denoting the displacement between vortices $i$ and $j$, and $r_{ij} = |\Vec{x}_{ij}|$. 
    The second terms are the induced velocities at vortex $i$ due to image vortices, which implement boundary conditions, with  $\Vec{x}_{ij, {\rm im}} = \Vec{x}_{i} - \Vec{x}_{j, {\rm im}} = (x_{ij, {\rm im}}, y_{ij, {\rm im}})$ denoting the displacement between vortex $i$ and the image of vortex $j$, and $r_{ij, {\rm im}} = |\Vec{x}_{ij, {\rm im}}|$\footnote{Following \citet{Howittetal2020}, we define $2\pi\kappa$ as the quantum of circulation in this appendix, rather than $\kappa$ as it is elsewhere in this paper.}.
    The third terms take into account the rotating reference frame, rotating with angular velocity $\Omega$, and comoving with the container. 
    The fourth term incorporates pinning, with $V(\Vec{x}_i - \Vec{x}_k)$ denoting the pinning potential at $\Vec{x}_i$ due to a pinning site at $\Vec{x}_k$. 
    $N_{\rm pin}$ denotes the total number of pinning sites.
    The pinning sites corotate with the container, so $\Vec{x}_k$ is fixed. 
    Pinning sites are arranged in a square lattice centred at the origin, and the pinning potential is an isotropic Gaussian,
    \begin{equation}
        V(r) = V_0 e^{-r^2/2\xi^2}~, \label{eq:pinningpotential}
    \end{equation}
    where $r$ is the distance from the pinning site to the vortex, and $\xi$ is the characteristic width of the pinning site \citep{Howittetal2020}. 
    The rotation matrix $\mathcal{R}_{\phi}$ in equation (\ref{eq:EOMall}) embeds the effect of the dissipation for viscous superfluids by rotating the vortex velocity by an angle $\phi$, as in \citet{Schwarz1985}; see also Section 2.6 of \citet{Howittetal2020}. 
    
    The superfluid communicates with the container through the angular momentum equation
    \begin{equation}
        \frac{\dd\Omega}{\dd t} = N_{\rm ext} - I_{\rm rel} \frac{\dd\Omega_{\rm s}}{\dd t}
        \label{eq:EOMangmom}
    \end{equation}
    where $I_{\rm rel} = I_{\rm s}/I_{\rm c}$ is the ratio between the superfluid moment of inertia $I_{\rm s}$ and container moment of inertia $I_{\rm c}$, and $N_{\rm ext}$ is the external spin-down torque divided by $I_{\rm c}$. 
    $\Omega_{\rm s}$ is the angular velocity of the superfluid, computed from the superfluid angular momentum $L_{\rm s} = I_{\rm s} \Omega_{\rm s}$, which in turn is related to the vortex configuration through
    \begin{equation}
        L_{\rm s} = \kappa \sum_{i=1}^{N_{\rm v}} \left[R_\ast^2 - (x_i^2 + y_i^2)\right]~.
    \end{equation}
    The equations of motion (\ref{eq:EOMall}) -- (\ref{eq:EOMangmom}) are solved numerically using the Runge-Kutta Cash-Karp (RKCK) scheme \citep{PressEtAl1992}. 
    
    The units of the simulations are set such that $\kappa = 1$, $R_\ast = 10.0$, and $I_{\rm rel} = 1.0$. 
    The spacing between pinning sites is $a_{\rm pin}=0.1$, the characteristic width of the pinning sites is $\xi = 0.02$, and the pinning strength is $V_0 = 2$. 
    To accelerate the code, we only consider the closest pinning site to a given vortex in equations (\ref{eq:EOMx}) and (\ref{eq:EOMy}). 
    The dissipation angle is set to be $\phi = 0.1$. 
    
    The simulations are initialized by drawing positions of $5\times10^3$ vortices from a spatially uniform probability distribution and evolving without spindown, until all the vortices are pinned. 
    During this evolution, $\lesssim 60$ vortices leave the system. 
    Then, we turn on the external spin-down torque $N_{\rm ext} = -2.5\times10^{-2}$ in simulation units.
    The time-step is $\delta t = 2\times10^{-3}$ in simulation units. 
    All simulations run for $2\times10^6$ time-steps or until the angular frequency of the container reaches zero, whichever comes first. 
    For each run, $\Vec{x}_1$, $\dots$, $\Vec{x}_{N_{\rm v}}$ and $\Omega$ are recorded every 50 time-steps to reduce the data volume. 
    Table \ref{tab:SimParams} includes a summary of control parameters for the simulation. For more details, see Sections 2 and 3 and Table 1 in \citet{Howittetal2020}. 

    \begin{table*}
        \centering
        \begin{tabular}{lcc}
            \hline
            Parameter & Value (simulation units) & Physical meaning \\
            \hline 
            $\kappa$ & 1 & Quantum of circulation$/(2\pi)$ \\
            $R_\ast$ & 10.0 & Radius of container \\
            $I_{\rm rel}$ & 1.0 & Ratio of superfluid/container moments of inertia \\
            $a_{\rm pin}$ & 0.1 & Spacing between pinning sites \\
            $\xi$ & $2\times10^{-2}$ & Characteristic width of pinning site \\
            $V_0$ & 2 & Pinning strength \\
            $N_{\rm ext}$ & $-2.5\times10^{-2}$ & External spin-down torque divided by container moment of inertia \\
            $\phi$ & 0.1 & Dissipation angle \\
            $N_{\rm v}(t=0)$ & $5\times10^3$ & Initial number of vortices \\
            $\delta t$ & $2\times10^{-3}$ & Time-step \\
            \hline
        \end{tabular}
        \caption{Summary of control parameters for the vortex avalanche simulations in Section \ref{sec:QScalingForSim}.}
        \label{tab:SimParams}
    \end{table*}
    
    The angular frequency time-series is smoothed with a top-hat function of width 500 time-steps to reduce fluctuations from vortex ``jiggling'' \citep{WarszawskiMelatos2011}. 
    Glitches are found with an algorithm which scans the smoothed $\Omega$ time-series for time-steps, when the incremental change in $\Omega$ switches sign.

\section{Analytic PDF of \texorpdfstring{$Q$}{Q} for a uniform Poisson configuration}
    \label{sec:AppxPoisson}
    In this appendix, we calculate the probability distribution of $Q$ defined in equation (\ref{eq:Q}), when the ${N_{\rm v}}$ vortex positions $(R_i, \phi_{i, 0})$ in the corotating frame are generated by a uniform Poisson point process. 
    
    Consider a random vector $\Vec{W}_i$ with Cartesian components $W_{i, x} = (R_i/R_\ast)(1-R^2_i/R_\ast^2)^{3/2}\cos\phi_{i, 0}$ and $W_{i, y} = (R_i/R_\ast)(1-R^2_i/R_\ast^2)^{3/2}\sin\phi_{i, 0}$. By referring to equation (\ref{eq:S}), we have $\Vec{S}=\sum_{i=1}^{N_{\rm v}} \Vec{W}_i$. Suppose that $\Vec{x}_i = (R_i, \phi_{i, 0})$ is drawn from a uniform distribution with equal probability density per unit area in the disk of radius $R_\ast$. Then, the mean $\langle \Vec{W} \rangle$ vanishes, and the variance is given by
    \begin{align}
        {\rm var} ~ W_x 
         =& \frac{1}{\pi R_\ast^2} \int_0^{2\pi} \dd\phi_{0}\, \int_0^{R_\ast} \dd R\,  R \left(1-\frac{R^2}{R_\ast^2}\right)^3 \frac{R^2}{R_\ast^2} \cos^2{\phi_0} \\
         =& \frac{1}{40}~.
    \end{align}
    Analogously, we have ${\rm var} ~ W_y = 1/40$. 
    
    Let us draw ${N_{\rm v}} \gg 1$ independent samples of $\Vec{W}$. The central limit theorem implies $\langle S_x \rangle = 0 = \langle S_y \rangle$, ${\rm var} ~ S_x = {N_{\rm v}}/40 = {\rm var} ~ S_y$,
    \begin{equation}
        p(S_x) = \left(\frac{20}{\pi {N_{\rm v}}}\right)^{1/2} \exp{\left(-\frac{20S_x^2}{{N_{\rm v}}}\right)}
    \end{equation}
    and hence 
    \begin{equation}
        p(Q) = \left( \frac{40 Q^2}{{N_{\rm v}}} \right)^{1/2} \exp{\left(-\frac{20Q^2}{{N_{\rm v}}}\right)} ~. \label{eq:Q_rayleigh_uniform_again}   
    \end{equation}
    Equation (\ref{eq:Q_rayleigh_uniform_again}) is the same as equation (\ref{eq:Q_rayleigh_uniform}) in Section \ref{sec:UniformPoisson}. 
    
    The mean of $Q$ calculated from equation (\ref{eq:Q_rayleigh_uniform_again}) is $\langle Q \rangle = (\pi {N_{\rm v}}/80)^{1/2} \propto {N_{\rm v}}^{1/2}$. 
    This proportionality makes sense intuitively. 
    Expanding the squares of the sum in equation (\ref{eq:Q}) gives three types of terms: ones proportional to $\cos^2 \phi_{i, 0}$, ones proportional to $\sin^2 \phi_{i, 0}$, and ones proportional to the cosine of the difference between $\phi_{i,0}$ for vortex pairs. 
    If there is no correlation between vortex positions, the cross-terms cancel on average. 
    The terms proportional to $\cos^2 \phi_{i, 0}$ and $\sin^2 \phi_{i, 0}$ sum to a factor of order 1. 
    There are $N_{\rm v}$ of these terms, yielding $Q^2 \propto {N_{\rm v}}$ on average. 
    
    This calculation is a more detailed version of the central-limit-theorem argument in Section 5 of \citet{Melatosetal2015}.

\section{\texorpdfstring{$Q$}{Q} for a regular array}
    \label{sec:AppxArray}
    In this Appendix, we derive $Q$ as a function of ${N_{\rm v}}$ for a general array of vortices placed at regular intervals, i.e.\ a two-dimensional Bravais lattice. 
    
    A two-dimensional Bravais lattice is defined by two primitive translation vectors ${\Vec{d}_1, \Vec{d}_2}$. The lattice points are located at $\Vec{R}_{n_1 n_2} = n_1 \Vec{d}_1 + n_2 \Vec{d}_2 + \Vec{c}$, where $n_1$ and $n_2$ are integers and $\Vec{c}$ is an arbitrary but constant shift within the unit cell. In this paper, we place vortices on lattice points which lie inside the disk of radius $R_\ast$. The vortex density, i.e.\ the number of vortices per unit area, can therefore be written as
    \begin{equation}
        n_{\rm v} (\Vec{x}) = \sum_{n_1=-\infty}^{\infty} \sum_{n_2=-\infty}^{\infty}  \delta^2 (\Vec{x} - \Vec{R}_{n_1 n_2})~.
        \label{eq:arraydensity}
    \end{equation}
    To calculate the components of $\Vec{S}$ defined in equation (\ref{eq:S}), consider the complex-valued integral 
    \begin{equation}
        \tilde{S} = \int \dd^2 \Vec{x}\,  n_{\rm v} (\Vec{x}) \frac{R}{R_\ast}\left(1-\frac{R^2}{R_\ast^2}\right)^{3/2} e^{{\rm i}\phi}  
        \label{eq:complexS}
    \end{equation}
    with $\Vec{x} = (R, \phi)$ symbolizing cylindrical coordinates as in Section \ref{sec:CurrentQuadMoment}. The integral is performed over the disc of radius $R_\ast$ centred at the origin. The real and imaginary parts of $\Tilde{S}$ correspond to the $x$- and $y$-components of $\Vec{S}$, respectively. One has $Q = |\tilde{S}|$ for a regular array defined by equation (\ref{eq:arraydensity}).
    
    Equation (\ref{eq:arraydensity}) can be rewritten in terms of its reciprocal lattice (or equivalently, a Fourier series), viz.
    \begin{equation}
        n_{\rm v} (\Vec{x}) = \sum_{m_1=-\infty}^{\infty} \sum_{m_2=-\infty}^{\infty} B_{m_1 m_2} e^{{\rm i} \Vec{k}_{m_1 m_2} \cdot \Vec{x}}
    \end{equation}
    where $\Vec{k}_{m_1 m_2} = m_1 \Vec{b}_1 + m_2 \Vec{b}_2$ is the reciprocal lattice position vector, $\Vec{b}_1$ and $\Vec{b}_2$ are the reciprocal primitive translation vectors, and $m_1$ and $m_2$ are integers. The Fourier amplitudes $B_{m_1 m_2}$ are given by 
    \begin{align}
        B_{m_1 m_2} =&\, \frac{1}{d_1 d_2} \int_0^{d_1} \dd y_1\, \int_0^{d_2} \dd y_2\,  n_{\rm v} \left(y_1 \frac{\Vec{d}_1}{d_1} + y_2 \frac{\Vec{d}_2}{d_2}\right) \nonumber \\
        &\times e^{-{\rm i} \Vec{k}_{m_1 m_2} \cdot \Vec{x}} \label{eq:FourierAmplitudes}
    \end{align}
    where $d_i = |\Vec{d}_i|$ for $i \in \{1, 2\}$.
    Substituting equation (\ref{eq:arraydensity}) into equation (\ref{eq:FourierAmplitudes}), we obtain 
    \begin{equation}
        B_{m_1 m_2} = \frac{1}{d_1 d_2} e^{ -{\rm i} \Vec{k}_{m_1 m_2} \cdot \Vec{c} }~. \label{eq:FourierSeries}
    \end{equation}
    Substituting equation (\ref{eq:FourierSeries}) into equation (\ref{eq:complexS}) gives
    \begin{align}
        \tilde{S} =\;& \frac{1}{d_1 d_2} \int_0^{R_\ast} \dd R \int_0^{2\pi} \dd \phi\, R \left(\frac{R}{R_\ast}\right)\left(1-\frac{R^2}{R_\ast^2}\right)^{3/2} e^{{\rm i}\phi}  \nonumber \\
        ~&\times \sum_{m_1=-\infty}^{\infty} \sum_{m_2=-\infty}^{\infty} e^{{\rm i} \Vec{k}_{m_1 m_2} \cdot (\Vec{x}-\Vec{c})} \label{eq:complexS2}
    \end{align}
    The exponential term can be expanded using the Jacobi-Anger identity,
    \begin{align}
        e^{{\rm i} \Vec{k}_{m_1 m_2} \cdot \Vec{x}} =\;& e^{{\rm i} k_{m_1 m_2} R \cos(\phi_{m_1, m_2} - \phi)} \\
        =\;&  \sum_{n = -\infty}^{\infty} {\rm i}^n J_n(k_{m_1 m_2}R) e^{{\rm i}n(\phi_{m_1 m_2} - \phi)} \label{eq:JacobiAnger}
    \end{align}
    where we write $\Vec{k}_{m_1 m_2} = k_{m_1 m_2}(\cos{\phi_{m_1 m_2}}, \sin{\phi_{m_1 m_2}})$, and $J_n(z)$ denotes the $n$-th order Bessel function of the first kind. 
    Substituting equation (\ref{eq:JacobiAnger}) into equation (\ref{eq:complexS2}) and integrating with respect to $\phi$, we find that every term but $n = 1$ vanishes. 
    The result is 
    \begin{align}
        \tilde{S} =\;& \frac{2\pi {\rm i}}{d_1 d_2} \sum_{m_1=-\infty}^{\infty} \sum_{m_2=-\infty}^{\infty} e^{{\rm i} \Vec{k}_{m_1 m_2} \cdot \Vec{c}} e^{{\rm i}\phi_{m_1 m_2}} \nonumber \\ 
        ~&\times \int_0^{R_\ast} \dd R\, R \left(\frac{R}{R_\ast}\right)\left(1-\frac{R^2}{R_\ast^2}\right)^{3/2} J_n(k_{m_1 m_2}R) 
    \end{align}
    and hence
    \begin{equation}
        \tilde{S} = \frac{6\pi {\rm i} R_\ast^2}{d_1 d_2} \sum_{m_1=-\infty}^{\infty} \sum_{m_2=-\infty}^{\infty}  \frac{j_3(k_{m_1 m_2}R_\ast)}{(k_{m_1 m_2}R_\ast)^2} e^{{\rm i} \Vec{k}_{m_1 m_2} \cdot \Vec{c}} e^{{\rm i}\phi_{m_1 m_2}}~, \label{eq:Stilde_final}
    \end{equation}
    where $j_n(z)$ is the spherical Bessel function of order $n$. Evaluating equation (\ref{eq:Stilde_final}) for a square array gives equation (\ref{eq:Stilde_array}).
    
    The sum in equation (\ref{eq:Stilde_final}) vanishes for $\Vec{c} = \Vec{0}$, because the summand is an odd function of $m_1$ and $m_2$. 
    This corresponds to no emission of gravitational waves. 
    Physically, the wave strain vanishes because $\Vec{c} = 0$ corresponds to a configuration in which a vortex at $(R_1, \phi_1)$ has a counterpart opposite the centre of the star at $(R_1, \pi+\phi_1)$. These vortex pairs cancel perfectly. 
    Needless to say, perfect cancellation (e.g.\ $\Vec{c} = 0$) is unlikely in a real neutron star, where one expects $0 < |\Vec{c}| < (d_1^2 + d_2^2)^{1/2}$ in general.  
    
    We now calculate the scaling of $Q$-versus-${N_{\rm v}}$. Let us take $d_1 \sim d_2 \sim d$ and hence
    \begin{equation}
    k_{m_1 m_2} \sim \frac{2\pi}{d} (m_1^2 + m_2^2)^{1/2}~.    
    \end{equation}
    If the number of vortices is large, with ${N_{\rm v}} \approx \pi R_\ast^2 / d^2 \gg 1$, and hence $k_{m_1 m_2} R_\ast \gg 1$ for all $(m_1, m_2)$, the spherical Bessel function is approximated by Hankel's expansion, whose leading term is
    \begin{equation}
        j_3(k_{m_1 m_2} R_\ast) \approx \frac{\cos(k_{m_1 m_2} R_\ast)}{k_{m_1 m_2} R_\ast}. \label{eq:Hankel}
    \end{equation}
    Upon substituting equation (\ref{eq:Hankel}) into (\ref{eq:Stilde_final}), we find 
    \begin{align}
        \tilde{S} =\;& \frac{3{\rm i}}{4\pi^2}\frac{d}{R_\ast} \sum_{m_1=-\infty}^{\infty} \sum_{m_2=-\infty}^{\infty} \frac{e^{{\rm i} (\Vec{k}_{m_1 m_2} \cdot \Vec{c} + \phi_{m_1 m_2})}}{(m_1^2 + m_2^2)^{3/2}} \nonumber \\
        &\times  \cos\left[\frac{2\pi R_\ast}{d} (m_1^2 + m_2^2)^{1/2}\right].
        \label{eq:Stilde}
    \end{align}
    The summand of equation (\ref{eq:Stilde}) depends on ${N_{\rm v}}$ only through the factor $R_\ast/d \sim {N_{\rm v}}^{1/2}$. 
    The dependence lies in the argument of the cosine, which does not contribute to the magnitude of $\tilde{S}$. 
    The summand decreases with $(m_1^2 + m_2^2)^{-3/2}$, so the sum evaluates to $\mathcal{O}(1)$ at most. 
    Both the real and complex parts of $\tilde{S}$ are then of order $\mathcal{O}(d/R_\ast) = \mathcal{O}({N_{\rm v}}^{-1/2})$, yielding $Q \propto {N_{\rm v}}^{-1/2}$. 
    The $Q$-versus-${N_{\rm v}}$ scaling is insensitive to array geometry, as long as the array is regular, and one has ${N_{\rm v}} \gg 1$.


\bsp	
\label{lastpage}
\end{document}